\documentclass{an} 
\usepackage{natbib}
\usepackage{graphicx}
\usepackage{times}
\usepackage{fancyhdr}
\newcommand{\mj}{$M_{\mathrm{J}}$}
\newcommand{\rj}{$R_{\mathrm{J}}$}
\newcommand{\me}{$M_{\oplus}$}
\newcommand{\aap}{A\&A}

\begin{document}

\title{Young Jupiters are Faint:  New Models of the Early Evolution of Giant Planets}

\author{J.J. Fortney\inst{1} \and M.S. Marley\inst{1} \and O. Hubickyj\inst{1,2} \and P. Bodenheimer\inst{2} \and J.J. Lissauer\inst{1}}
\institute{Space Science and Astrobiology Division, NASA Ames Research Center, MS 245-3, Moffett Field, CA 94035
\and
UCO/Lick Observatory, University of California at Santa Cruz, Santa Cruz, CA 95064 }

\date{Submitted: 30 September, 2005}

\abstract{Here we show preliminary calculations of the cooling and contraction of a 2 \mj~planet.    These calculations, which are being extended to 1--10 \mj, differ from other published ``cooling tracks'' in that they include a core accretion-gas capture formation scenario, the leading theory for the formation of gas giant planets.  We find that the initial post-accretionary intrinsic luminosity of the planet is $\sim$3 times less than previously published models which use arbitrary initial conditions.  These differences last a few tens of millions of years.  Young giant planets are intrinsically fainter than has been previously appreciated.  We also discuss how uncertainties in atmospheric chemistry and the duration of the formation time of giant planets lead to challenges in deriving planetary physical properties from comparison with tabulated model values.
}

\correspondence{jfortney@arc.nasa.gov}

\maketitle

\section{Introduction}
The direct imaging of extrasolar planets around other stars has long been one of the ``holy grails'' of astronomy.  Now, ten years after the disovery of the first extrasolar giant planet (EGP), this era is apparently now at hand.  The two major discoveries to date are those of \citet{Chauvin04,Chauvin05}, who discovered an object of $5\pm 2$ \mj~in orbit around a 25 \mj~brown dwarf (2M1207), and \citet{Neu05}, who discovered a $\sim$1 -- 42 \mj~companion to the T Tauri star GQ Lup. Both of these systems are quite young, with ages estimated to be $\sim$10 Myr for 2M1207 and $\sim$2 Myr for GQ Lup.

It must be stressed, however, that these mass estimates are based on thermal evolution models, which are most suspect at young ages.  (This is readily apparent from the GQ Lup b mass error bar.)  Indeed, the most common evolution models for these brown dwarfs and planetary-mass objects \citep{Burrows97,Chabrier00b,Baraffe03} use an arbitrary initial condition---the planet is assumed to be fully convective with an initial arbitrarily large radius.  This is important because evolution models (or ``cooling tracks'') are used to assign masses for these dim objects.  While giant planets cool quickly and forget their initial conditions at ages of billions of years, at ages of $\sim$50 Myr or less, these initial conditions are important.  Using evolutionary models that do not include a valid formation mechanism to assign a mass to a young object is quite inaccurate, especially at the extremely young ages of these new planetary candidates.  Modelers have stressed this point.

This very issue has arisen in the past few years with the study of very young brown dwarfs.  \citet{Baraffe02} have analyzed various compilations of evolutionary models of low mass stars and brown dwarfs.  They also tested various initial conditions on their own models and find that the cooling tracks differ significantly at young ages (less than $\sim$5 Myr), which could lead to very different mass or age estimates, based on the luminosity of a given object, or group of objects.  \citet{Wuchterl03} have published formation models of stars and brown dwarfs down to 50 \mj.  These models follow the collapse of an initial cloud all the way to the main sequence, while making the minimum number of a priori assumptions.  These studies highlight the importance of initial conditions to the formation of stars and brown dwarfs.

In tandem with this theoretical work, new observations have begun to test the predictions of evolution models at young ages.  These include \citet{Close05} and \citet{Reiners05} who, with dynamical mass determinations and age estimates, find that commonly used M dwarf evolutionary tracks \citep[e.g.][]{Burrows97,Chabrier00b} overestimate the luminosity of these objects at ages of $\sim$50 -- 100 Myr (for the \citeauthor{Close05} work) and $\sim$5 Myr (for the \citeauthor{Reiners05} work).  \citet{Mohanty04a,Mohanty04b} have used high resolution synthetic spectra to determine $T_{\mathrm{eff}}$ and gravity for a variety of objects down to perhaps 10 \mj~in Upper Scorpius, and find disagreement with the \citet{Chabrier00b} models.  Reiners et al.~, in this volume, update the Mohanty et al.~work with new model spectra and find that models are consistently overestimating the luminosity at young ages.  Therefore, as of July 2005, there appears to be disagreement between these observations and models (although \citealt{Luhman05} question the \citeauthor{Close05} age estimate for the object AB Dor).

To date, there has been little attention paid to the dependence on formation to the evolutionary history of planets.  The two leading scenarios for the formation of giant planets are the core accretion-gas capture theory \citep[e.g.,][]{Pollack96}, which is the generally accepted theory of giant planet formation, and the direct gravitational collapse theory \citep[e.g.,][]{Boss01,Mayer04}, which has some proponents.  Here we investigate the evolution of a 2 \mj~giant planet, under the assumption that the planet forms via the core accretion-gas capture method.  We compare the evolutionary tracks to those that assume what we term an ``initially hot start,'' meaning the planets are assumed to be adiabatic at every age.  For these types of models it is common practice to pick an initially very hot adiabat, from which point the planet cools extremely rapidly, and then plot the cooling history for reasonable ages, say from 1 Myr to 10 Gyr.  These models are typified by \citet{Burrows97}, \citet{Baraffe03}, and \citet{FH04}.  Our models that incorporate the core accretion-gas capture formation mechanism are considerably less luminous at  young ages than other published models, and these differences can last tens of millions of years, not just ``a few'' million years.

\section{Methods}
\subsection{Core Accretion-Gas Capture Model}
\begin{figure}
\resizebox{\hsize}{!}
{\includegraphics[]{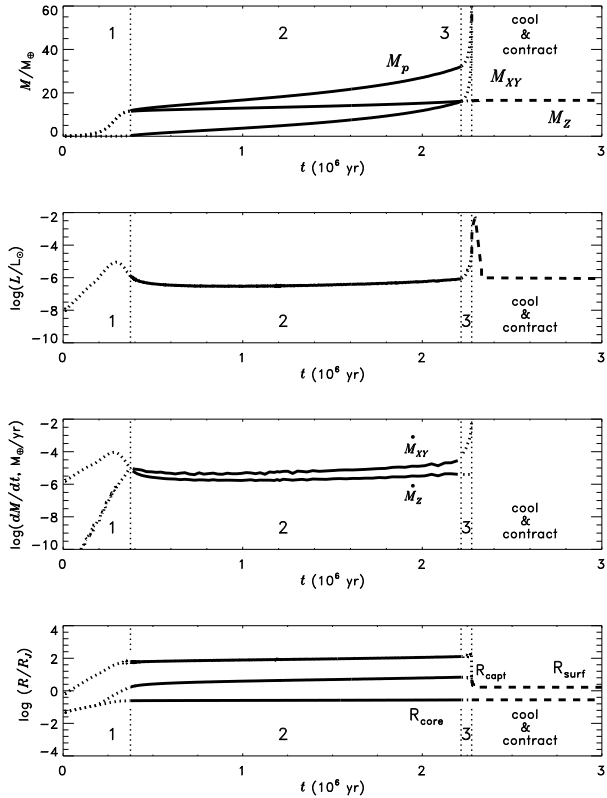}}
\caption{The mass (units of M$_{\oplus}$), the luminosity (units of L$_{\odot}$), the accretion rates (units of M$_{\oplus}$/year), and the radii (units of \rj) are plotted as a function of time (units of million years) for the \citet{Hubickyj05} baseline formation model of a 1 \mj~planet.  The first three phases of formation of the protoplanet (see text) are labeled. {\it Dotted line:} phase 1, {\it solid line:} phase 2, {\it second dotted line}: phase 3. The dashed line is contraction and cooling (and is merely illustrative here), at which time we switch to an evolution code with fully nongray atmosphere models.}
\label{fig1}
\end{figure}

The core accretion-gas capture model (a.k.a. core nucleated accretion) of giant planet formation \citep{Mizuno80,Bodenheimer86,Pollack96} involves the formation of a giant planet by capture of nebula gas by a solid core. These stages are as follows \citep{Bodenheimer00}:

\begin{enumerate}
\item Dust particles in the solar nebula form planetesimals that accrete into a solid core surrounded by a low mass gaseous envelope. Initially, solid runaway accretion occurs, and the gas accretion rate is much lower than that of solids. As the solid material in the feeding zone is depleted, the solid accretion rate is reduced.
\item The gas accretion rate steadily increases and eventually exceeds the solid accretion rate. The protoplanet continues to grow as the gas accretes at a relatively constant rate. The mass of the solid core also increases but at a slower rate, and eventually the core and envelope masses become equal.
\item Runaway gas accretion occurs and the protoplanet grows rapidly. The evolution up to this point is referrred to as the nebular stage, because the outer boundary of the protoplanetary envelope is in contact with the solar nebula and the density and temperature at this interface are given nebular values. (See Figure \ref{fig1}.)
\item The gas accretion rate reaches a limiting value defined by the rate at which the nebula can transport gas to the vicinity of the planet. Subsequent to this point, the equilibrium region of the protoplanet contracts inside the effective accretion radius, and gas accretes hydrodynamically onto this equilibrium region. This part of the evolution is considered to be the transition stage.
\item Accretion is stopped by either the opening of a gap in the disk as a consequence of accretion and the tidal effect of the planet, or by dissipation of the nebula. Once accretion stops, the planet enters the isolation stage.
\item The planet contracts and cools to the present state at constant mass. (It is the earliest part of stage 6 that is a focus of this article.)
\end{enumerate}

Simulations of this process have been successful in explaining many of the observed characteristics of the giant planets \citep{Pollack96}, although some important unresolved issues remain. Perhaps chief among these is the abundance of atmospheric dust grains, which play a major role in controlling the timescale of accretion. Recent calculations \citep{Hubickyj05} show that a model in which interstellar medium (ISM)-like grains are found throughout the accreting protoplanetary envelope takes 6 Myr before rapid gas accretion onto the planetary core begins. A comparable model in which the grain opacity is arbitrarily reduced to 2\% of the ISM value, presumably because of sedimentation of grains from high in the atmosphere to their condensation level, takes only 2.2 Myr to reach this point. The physical mechanism involved is that an atmosphere with grains forms a more effective thermal blanket so that the planet retains heat from its accretion for a longer time. The planet takes longer to contract and thus the onset of rapid gas accretion is delayed.

Recent calculations by \citet{Podolak03} indicate that these low opacity cases are likely closer to reality, and the actual opacity may well be less than the ``2\% of ISM'' used by \citet{Hubickyj05}. This conclusion is also supported by observations of cloudy L-dwarfs, where models that assume the opacity of grains as given by the ISM distribution, without sedimentation, are a very poor match to the data \citep{Chabrier00b} because they greatly overestimate the dust opacity. However, models that allow for grains to grow and sediment (see \citealp{AM01}) match spectroscopic data very well \citep{Marley02,Burgasser02,Marley04}. In short, the timescale argument against the core accretion-gas capture scenario is much weaker than had been previously thought, when opacities more appropriate to protoplanetary atmosphere conditions are used.

\subsection{Initial Conditions for Cooling}
Uncertainties in the opacity of the post-formation atmosphere, when the ``isolated'' planet begins to cool, lead to uncertainties in the luminosity, spectra, and colors of a newly formed planet. Although the planet formation code could be used to follow the subsequent cooling of the planets, the grain-laden atmospheres that are assumed, while perhaps realistic during the formation phase when many Earth masses of solid planetesimals are ablating into the proto-atmosphere, are not correct post-formation. After formation, our procedure is to switch over to our fully non-gray EGP/brown dwarf atmosphere code, so that we can compute a realistic atmosphere grid for the planets' subsequent evolution.  This evolution is computed with a separate planetary evolution code \citep[see][]{FH03}.  We are also able to use our atmosphere code to compute spectra and infrared colors, which the planetary formation code cannot do.

Specifically, our procedure is to take the radius (at a pressure of 1 bar) of the fully formed planet, just after accretion has terminated, and use this as the starting point for subsequent evolution.  The first model planet calculated with the structure/evolution code must match this same radius.  (Essentially no change in the internal structure accompanies this change in models as both codes use the \citealt{SCVH} H/He equation of state.)  The planet then contracts and cools over time, as the adiabatic interior cools and the atmosphere radiates this energy away.  These models include a 3.2 g cm$^{-3}$ solid core which grows during formation and reaches 16 \me~by the time the planet is fully formed.  The exact structure of the core has little effect on the evolution, although in the future it will be worthwhile to inlude the effects of self-compression on the core density.

\subsection{Atmosphere Models}
We have computed a fully nongray grid of model atmospheres, which serve as the upper boundary condition in the evolutionary calculation.  This grid spans a range in $T_{\mathrm{eff}}$ from 85 to 1500 K and in gravity from log $g$ = 2 to 4 (where $g$ is in cm s$^{-2}$).  To compute our grid we employ a model atmosphere code that has been used for a variety of planetary and substellar objects.  The code was first used to generate pressure-temperature (\emph{P--T}) profiles and spectra for Titan by \citet{Mckay89}.  It was significantly revised to model the atmospheres of brown dwarfs \citep{Marley96, Burrows97, Marley02}, Uranus \citep{MM99}, cool EGPs \citep{Marley98}, and ``Hot Jupiters'' \citep{Fortney05}.  The basic radiative transfer solving scheme was developed by \citet{Toon89}.

We use the elemental abundance data of \citet{Lodders03} and compute chemical equilibrium compositions following \citet{Fegley94} and \citet{Lodders02}.  In addition, we maintain a large and constantly updated opacity database.  Each individual radiative-convective \emph{P--T} profile is arrived at iteratively until the net flux in each layer is conserved to at least one part in $10^6$.  Here we neglect the opacity of silicate and iron clouds, as this study is still somewhat provisional.  The models that we compare to, \citet{Burrows97} and \citet{Baraffe03}, also do not include the opacity of clouds.

\section{Planetary Evolution}
The main conclusion of this work is shown in Fig.~2.  Shown is the luminosity of a 2 \mj~planet from an age of 10$^5$ to 10$^9$ years.  The models shown are:  thick solid curve (this work), dotted curve (\citealt{Burrows97}), dashed curve (\citealt{Baraffe03}), and thin solid curve (a variant on our work, to be discussed in Section 4.2).  Post-formation, at an age of 2.5 Myr, our models predict a luminosity only $\sim$1/3 of that predicted by the Burrows et al.~and Baraffe et al.~hot-start models.  This difference remains large for tens of millions of years, and Burrows/Baraffe models are still $\sim$50\% more luminous at an age of 20 Myr.  Gradually, the initial conditions of the various models are forgotten, and cooling tracks run together.  This is expected, since all of these calculations use the same H/He equation of state \citep{SCVH} and fairly similar atmosphere models.  This is a representative calculation, but our preliminary calculations suggest somewhat larger differences at larger masses.  However, this will need to be investigated further.

\begin{figure}
\resizebox{\hsize}{!}
{\includegraphics[]{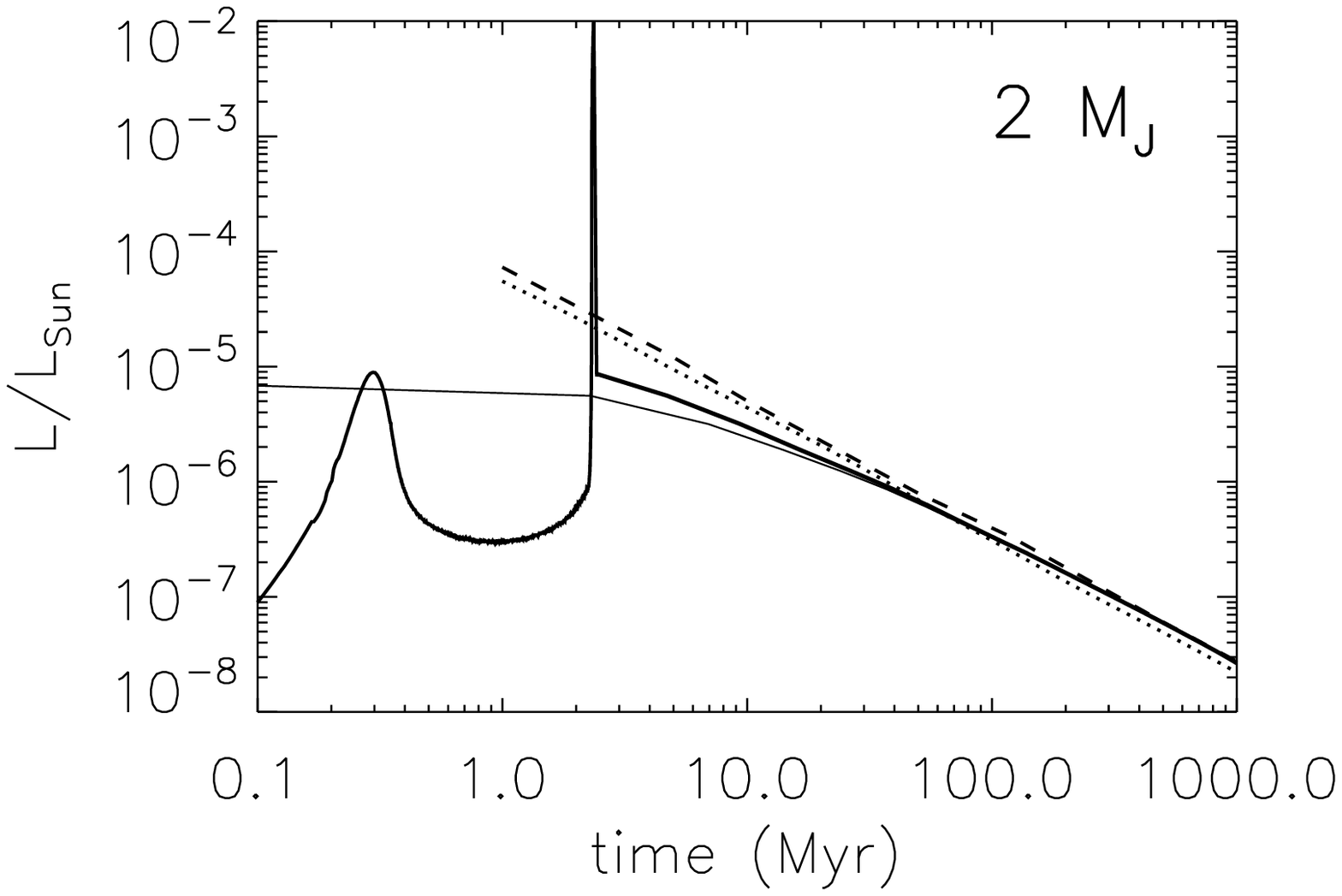}}
\caption{Luminosity vs.~time for a 2 \mj~planet for three sets of models.  The thick solid curve shows our new calculation, including the core accretion-gas capture formation mechanism.  The planet is fully formed at 2.2 Myr.  The dotted curve is for the models of \citet{Burrows97}.  The dashed curve is for the models of \citet{Baraffe03}.  The thin solid line is also our calculation, but the first-post accretion model planet is placed at $t$ = 0.  The full-width at half-maximum of the accretion luminosity (\emph{not} the log of the luminosity) spike is $\sim$40,000 years.  In reality, the spike is likely to be less sharp because of gradual accretion across the gap that the protoplanet  forms.} 
\label{fig2}
\end{figure} 

The implications of this calculation for the discovery of young Jupiter-mass planets is clear.  Since hot-start models systematically overestimate the luminosity of young giant planets, deriving planetary masses based on comparisons with cooling tracks, using an object's assumed age and observed luminosity, will lead one to systematically \emph{underestimate} the true masses of the planets.

\section{Discussion}
\subsection{Evolution Models}
The implication of our results is that giant planets formed by the core accretion-gas capture mechanism are less luminous post-accretion than had been previously appreciated.  A signficant amount of energy is radiated away in the formation process.   The fully formed planet has a smaller radius at young ages than hot start models predict, leading to a lower post-formation luminosity.  Prior evolutionary model tabulations overestimate the luminosity of giant planets at young ages.  The initial conditions for evolution models are not ``forgotten'' for tens of millions of years.  Thus, these planets will be more difficult to detect at young ages. 

It is illustrative to compare the integrated radiated energy (luminosity $\times$ time) of the various models as function of age.  This is shown in Fig.~3.  The line styles are the same as those in Fig.~2.  The location of the luminosity spike at 2.2 Myr is obvious for our calculation.  The \citet{Baraffe03} energy curve is always higher than the \citet{Burrows97} curve because as seen in Fig.~2, the luminosity of the \citeauthor{Baraffe03} model is larger at every age.  The integrated luminosity of our new model is greater than either of these models.  However, if we take our own ``hot start'' model (dash-dot curve), which is very similar to Baraffe/Burrows, and start it at an even younger age of 0.1 Myr, the integrated radiated energy at 1 Gyr approches that of our core accretion-gas capture model.  If the starting age for our hot start model is pushed back even younger, to 0.01 Myr (not shown), the integrated energy at 1 Gyr exceeds that of the core accretion-gas capture model.  (It is unlikely that adiabatic hot start models have relevance at these extremely young ages.  However, this shows the sensitiviy of the integrated radiated luminosity to the initial age for these models.)  The integrated radiative energies are all in the range expected, as we find that the gravitational binding energy of a 2 \mj~planet at 1 Gyr is -8.8$\times 10^{43}$ ergs.

\begin{figure}
\resizebox{\hsize}{!}
{\includegraphics[]{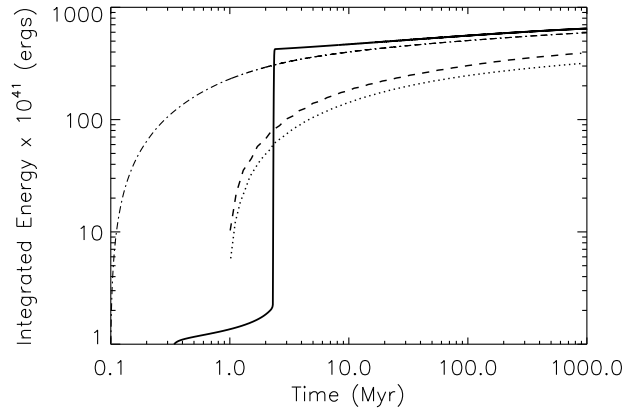}}
\caption{Integrated radiated energy vs.~time for a 2 \mj~planet.  As in Fig.~2, the thick solid curve shows our new calculation, including the core accretion-gas capture formation mechanism.  The dotted curve is for the models of \citet{Burrows97}.  The dashed curve is for the models of \citet{Baraffe03}.  The dash-dot line is for our own ``hot-start'' models, which are very similar to those are Burrows/Baraffe, but here we integrate back to 10 times younger ages (0.1 Myr).}
\label{fig2}
\end{figure} 

\subsection{Other Detectability Issues}
Another factor that may lead to young Jupiter-mass planets being dimmer than previously expected is the result of non-equilibrium carbon chemistry.  It was first suggested by \citet{Marley96}, after the discovery of Gl229b, that a substantial M-band 4 - 5 $\mu$m flux peak should be a universal feature of giant planets and brown dwarfs.  In addition to the intrinsic emergent flux, this spectral range has looked promising for planet detection due to the favorable planet/star flux ratio.  However, it has been known since the 1970s \citep[see][]{Prinn77} that the 5 $\mu$m flux in Jupiter is less than one predicts from a chemical equilibrium calculation.  This is due the dredging up of CO from deeper layers in Jupiter's atmosphere.  The chemical conversion timescale to CH$_4$ is much slower than the mixing timescale, ensuring a mixing ratio of CO several orders of magnitude above the equilibrium mixing ratio.  This same effect has now been observed in brown dwarf M-band photometry \citep{Golim04}, as anticipated by \citet{Fegley96}.  This excess CO leads to strong absorption at 4.5 $\mu$m, leading to diminished flux in M-band \citep{Saumon03}.  Consequently, when searching for giant planets, L-band may be just as favorable as M-band.

A factor that seemingly has been ignored to date in the detection of young objects around other stars is the actual formation time of the planetary candidates.  Since the formation timescale of giant planets via core accretion-gas capture is $\sim$1 -- 10 Myr, the assumption of co-evality with the parent star, which is used for assigning an age to the planet, is rather dubious.  Interestingly, this effect will work in reverse to the intrinsic faintness of the young giant planet---these giant planets will be younger than one would assume from co-evality---meaning a less massive planet could produce this same flux at the given stellar age.  It is worth remembering that the planetary formation timescale is uncertain, and that assuming co-evality at ages of only a few Myr could lead to significant errors in determining the physical properties of young objects.  We note that the discrepancy between our new models and those of \citeauthor{Burrows97} and \citeauthor{Baraffe03} would be even greater if our first post-formation planet (here age 2.2 Myr) was assigned an age of 0 Myr.  This is shown as the thin solid line in Fig.~2.  The luminosity vs.~time curve is shifted to the left considerably.  This further highlights the uncertainties in applying evolution models at young ages.

Finally, we note that given the large physical separations of the objects discovered by \citet{Chauvin04} and \citet{Neu05} from their parent stars, it is unlikely that these particular objects formed via the core accretion-gas capture method.  This indeed causes some to claim these objects are not ``planets'' at all, which remains a point of active discussion.  Whatever the formation mechanism for these objects, the point remains that evolution models are uncertain at young ages, and that new models and further observational calibrations are needed.  Our work shows that young Jupiters are fainter than has been previously appreciated.  In tandem with exciting observations, our knowledge about low-mass objects at young ages continues to grow.

\acknowledgements
JJF is supported by a postdoctoral research fellowship from the National Research Council (NRC) of the United States.  MSM, OH, \& PB are supported by grants from the NASA Origins program.  JJL is supported by a grant from the NASA Outer Planets program.


\begin{thebibliography}{41}
\expandafter\ifx\csname natexlab\endcsname\relax\def\natexlab#1{#1}\fi

\bibitem[{{Ackerman} \& {Marley}(2001)}]{AM01}
{Ackerman}, A.~S. \& {Marley}, M.~S. 2001, \apj, 556, 872

\bibitem[{{Baraffe} {et~al.}(2002){Baraffe}, {Chabrier}, {Allard}, \&
  {Hauschildt}}]{Baraffe02}
{Baraffe}, I., {Chabrier}, G., {Allard}, F., \& {Hauschildt}, P.~H. 2002, A\&A,
  382, 563

\bibitem[{{Baraffe} {et~al.}(2003){Baraffe}, {Chabrier}, {Barman}, {Allard}, \&
  {Hauschildt}}]{Baraffe03}
{Baraffe}, I., {Chabrier}, G., {Barman}, T.~S., {Allard}, F., \& {Hauschildt},
  P.~H. 2003, \aap, 402, 701

\bibitem[{{Bodenheimer} {et~al.}(2000){Bodenheimer}, {Hubickyj}, \&
  {Lissauer}}]{Bodenheimer00}
{Bodenheimer}, P., {Hubickyj}, O., \& {Lissauer}, J.~J. 2000, Icarus, 143, 2

\bibitem[{{Bodenheimer} \& {Pollack}(1986)}]{Bodenheimer86}
{Bodenheimer}, P. \& {Pollack}, J.~B. 1986, Icarus, 67, 391

\bibitem[{{Boss}(2001)}]{Boss01}
{Boss}, A.~P. 2001, \apj, 563, 367

\bibitem[{{Burgasser} {et~al.}(2002){Burgasser}, {Marley}, {Ackerman},
  {Saumon}, {Lodders}, {Dahn}, {Harris}, \& {Kirkpatrick}}]{Burgasser02}
{Burgasser}, A.~J., {Marley}, M.~S., {Ackerman}, A.~S., {Saumon}, D.,
  {Lodders}, K., {Dahn}, C.~C., {Harris}, H.~C., \& {Kirkpatrick}, J.~D. 2002,
  \apjl, 571, L151

\bibitem[{{Burrows} {et~al.}(1997){Burrows}, {Marley}, {Hubbard}, {Lunine},
  {Guillot}, {Saumon}, {Freedman}, {Sudarsky}, \& {Sharp}}]{Burrows97}
{Burrows}, A., {Marley}, M., {Hubbard}, W.~B., {Lunine}, J.~I., {Guillot}, T.,
  {Saumon}, D., {Freedman}, R., {Sudarsky}, D., \& {Sharp}, C. 1997, \apj, 491,
  856

\bibitem[{{Chabrier} {et~al.}(2000){Chabrier}, {Baraffe}, {Allard}, \&
  {Hauschildt}}]{Chabrier00b}
{Chabrier}, G., {Baraffe}, I., {Allard}, F., \& {Hauschildt}, P. 2000, \apj,
  542, 464

\bibitem[{{Chauvin} {et~al.}(2004){Chauvin}, {Lagrange}, {Dumas}, {Zuckerman},
  {Mouillet}, {Song}, {Beuzit}, \& {Lowrance}}]{Chauvin04}
{Chauvin}, G., {Lagrange}, A.-M., {Dumas}, C., {Zuckerman}, B., {Mouillet}, D.,
  {Song}, I., {Beuzit}, J.-L., \& {Lowrance}, P. 2004, A\&A, 425, L29

\bibitem[{{Chauvin} {et~al.}(2005){Chauvin}, {Lagrange}, {Dumas}, {Zuckerman},
  {Mouillet}, {Song}, {Beuzit}, \& {Lowrance}}]{Chauvin05}
---. 2005, A\&A, 438, L25

\bibitem[{{Close} {et~al.}(2005){Close}, {Lenzen}, {Guirado}, {Nielsen},
  {Mamajek}, {Brandner}, {Hartung}, {Lidman}, \& {Biller}}]{Close05}
{Close}, L.~M., {Lenzen}, R., {Guirado}, J.~C., {Nielsen}, E.~L., {Mamajek},
  E.~E., {Brandner}, W., {Hartung}, M., {Lidman}, C., \& {Biller}, B. 2005,
  Nature, 433, 286

\bibitem[{{Fegley} \& {Lodders}(1994)}]{Fegley94}
{Fegley}, B.~J. \& {Lodders}, K. 1994, Icarus, 110, 117

\bibitem[{{Fegley} \& {Lodders}(1996)}]{Fegley96}
---. 1996, \apjl, 472, L37

\bibitem[{{Fortney} \& {Hubbard}(2003)}]{FH03}
{Fortney}, J.~J. \& {Hubbard}, W.~B. 2003, Icarus, 164, 228

\bibitem[{{Fortney} \& {Hubbard}(2004)}]{FH04}
---. 2004, \apj, 608, 1039

\bibitem[{{Fortney} {et~al.}(2005){Fortney}, {Marley}, {Lodders}, {Saumon}, \&
  {Freedman}}]{Fortney05}
{Fortney}, J.~J., {Marley}, M.~S., {Lodders}, K., {Saumon}, D., \& {Freedman},
  R. 2005, \apjl, 627, L69

\bibitem[{{Golimowski} {et~al.}(2004){Golimowski}, {Leggett}, {Marley}, {Fan},
  {Geballe}, {Knapp}, {Vrba}, {Henden}, {Luginbuhl}, {Guetter}, {Munn},
  {Canzian}, {Zheng}, {Tsvetanov}, {Chiu}, {Glazebrook}, {Hoversten},
  {Schneider}, \& {Brinkmann}}]{Golim04}
{Golimowski}, D.~A., {Leggett}, S.~K., {Marley}, M.~S., {Fan}, X., {Geballe},
  T.~R., {Knapp}, G.~R., {Vrba}, F.~J., {Henden}, A.~A., {Luginbuhl}, C.~B.,
  {Guetter}, H.~H., {Munn}, J.~A., {Canzian}, B., {Zheng}, W., {Tsvetanov},
  Z.~I., {Chiu}, K., {Glazebrook}, K., {Hoversten}, E.~A., {Schneider}, D.~P.,
  \& {Brinkmann}, J. 2004, \aj, 127, 3516

\bibitem[{{Hubickyj} {et~al.}(2005){Hubickyj}, {Bodenheimer}, \&
  {Lissauer}}]{Hubickyj05}
{Hubickyj}, O., {Bodenheimer}, P., \& {Lissauer}, J.~J. 2005, Icarus, in press

\bibitem[{{Lodders}(2003)}]{Lodders03}
{Lodders}, K. 2003, \apj, 591, 1220

\bibitem[{{Lodders} \& {Fegley}(2002)}]{Lodders02}
{Lodders}, K. \& {Fegley}, B. 2002, Icarus, 155, 393

\bibitem[{{Luhman} {et~al.}(2005){Luhman}, {Stauffer}, \& {Mamajek}}]{Luhman05}
{Luhman}, K.~L., {Stauffer}, J.~R., \& {Mamajek}, E.~E. 2005, \apjl, 628, L69

\bibitem[{{Marley}(1998)}]{Marley98}
{Marley}, M.~S. 1998, in in ASP Conf. Ser. 134, Brown Dwarfs and Extrasolar
  Planets, ed. R. Rebolo, E. L. Martin, \& M. R. Zapatero Osorio (San
  Francisco: ASP), 383

\bibitem[{{Marley} {et~al.}(2004){Marley}, {Cushing}, \& {Saumon}}]{Marley04}
{Marley}, M.~S., {Cushing}, M.~C., \& {Saumon}, D. 2004, 13th Workshop on Cool
  Stars, Stellar Systems, and the Sun, eds. F.~Favata, G.~A.~J.~Hussain, \&
  B.~Battrick, ESA SP-560, p.794

\bibitem[{{Marley} \& {McKay}(1999)}]{MM99}
{Marley}, M.~S. \& {McKay}, C.~P. 1999, Icarus, 138, 268

\bibitem[{{Marley} {et~al.}(1996){Marley}, {Saumon}, {Guillot}, {Freedman},
  {Hubbard}, {Burrows}, \& {Lunine}}]{Marley96}
{Marley}, M.~S., {Saumon}, D., {Guillot}, T., {Freedman}, R.~S., {Hubbard},
  W.~B., {Burrows}, A., \& {Lunine}, J.~I. 1996, Science, 272, 1919

\bibitem[{{Marley} {et~al.}(2002){Marley}, {Seager}, {Saumon}, {Lodders},
  {Ackerman}, {Freedman}, \& {Fan}}]{Marley02}
{Marley}, M.~S., {Seager}, S., {Saumon}, D., {Lodders}, K., {Ackerman}, A.~S.,
  {Freedman}, R.~S., \& {Fan}, X. 2002, \apj, 568, 335

\bibitem[{{Mayer} {et~al.}(2004){Mayer}, {Quinn}, {Wadsley}, \&
  {Stadel}}]{Mayer04}
{Mayer}, L., {Quinn}, T., {Wadsley}, J., \& {Stadel}, J. 2004, \apj, 609, 1045

\bibitem[{{McKay} {et~al.}(1989){McKay}, {Pollack}, \& {Courtin}}]{Mckay89}
{McKay}, C.~P., {Pollack}, J.~B., \& {Courtin}, R. 1989, Icarus, 80, 23

\bibitem[{{Mizuno}(1980)}]{Mizuno80}
{Mizuno}, H. 1980, Progress of Theoretical Physics, 64, 544

\bibitem[{{Mohanty} {et~al.}(2004{\natexlab{a}}){Mohanty}, {Basri},
  {Jayawardhana}, {Allard}, {Hauschildt}, \& {Ardila}}]{Mohanty04a}
{Mohanty}, S., {Basri}, G., {Jayawardhana}, R., {Allard}, F., {Hauschildt}, P.,
  \& {Ardila}, D. 2004{\natexlab{a}}, \apj, 609, 854

\bibitem[{{Mohanty} {et~al.}(2004{\natexlab{b}}){Mohanty}, {Jayawardhana}, \&
  {Basri}}]{Mohanty04b}
{Mohanty}, S., {Jayawardhana}, R., \& {Basri}, G. 2004{\natexlab{b}}, \apj,
  609, 885

\bibitem[{{Neuh{\" a}user} {et~al.}(2005){Neuh{\" a}user}, {Guenther},
  {Wuchterl}, {Mugrauer}, {Bedalov}, \& {Hauschildt}}]{Neu05}
{Neuh{\" a}user}, R., {Guenther}, E.~W., {Wuchterl}, G., {Mugrauer}, M.,
  {Bedalov}, A., \& {Hauschildt}, P.~H. 2005, A\&A, 435, L13

\bibitem[{{Podolak}(2003)}]{Podolak03}
{Podolak}, M. 2003, Icarus, 165, 428

\bibitem[{{Pollack} {et~al.}(1996){Pollack}, {Hubickyj}, {Bodenheimer},
  {Lissauer}, {Podolak}, \& {Greenzweig}}]{Pollack96}
{Pollack}, J.~B., {Hubickyj}, O., {Bodenheimer}, P., {Lissauer}, J.~J.,
  {Podolak}, M., \& {Greenzweig}, Y. 1996, Icarus, 124, 62

\bibitem[{{Prinn} \& {Barshay}(1977)}]{Prinn77}
{Prinn}, R.~G. \& {Barshay}, S.~S. 1977, Science, 198, 1031

\bibitem[{{Reiners} {et~al.}(2005){Reiners}, {Basri}, \& {Mohanty}}]{Reiners05}
{Reiners}, A., {Basri}, G., \& {Mohanty}, S. 2005, ArXiv Astrophysics e-prints,
  astro-ph/0506501, ApJ in press

\bibitem[{{Saumon} {et~al.}(1995){Saumon}, {Chabrier}, \& {van Horn}}]{SCVH}
{Saumon}, D., {Chabrier}, G., \& {van Horn}, H.~M. 1995, \apjs, 99, 713

\bibitem[{{Saumon} {et~al.}(2003){Saumon}, {Marley}, {Lodders}, \&
  {Freedman}}]{Saumon03}
{Saumon}, D., {Marley}, M.~S., {Lodders}, K., \& {Freedman}, R.~S. 2003, in IAU
  Symposium 211: Brown Dwarfs, 345

\bibitem[{{Toon} {et~al.}(1989){Toon}, {McKay}, {Ackerman}, \&
  {Santhanam}}]{Toon89}
{Toon}, O.~B., {McKay}, C.~P., {Ackerman}, T.~P., \& {Santhanam}, K. 1989,
  Journal of Geophysical Research, 94, 16287

\bibitem[{{Wuchterl} \& {Tscharnuter}(2003)}]{Wuchterl03}
{Wuchterl}, G. \& {Tscharnuter}, W.~M. 2003, A\&A, 398, 1081

\end{thebibliography}



\end{document}